 \documentclass[final,3p,times]{elsarticle-mod}
\usepackage{amssymb,amsmath,bm}
\begin{document}

\renewcommand{\labelenumi}{(\roman{enumi})}

\begin{frontmatter}

\title{Two variants of the MCV3 scheme}

\author[titech]{Feng Xiao }\corref{cor}
%\author[uosaka]{Satoshi Ii}
%\author[xju]{Chungang Chen}
%\author[cma]{Xingliang Li}

\address[titech]{Department of energy sciences, Tokyo Institute of Technology,   
4259 Nagatsuta, Midori-ku, Yokohama, 226-8502, Japan}
%\address[uosaka]{ Department of Mechanical Science \& Bioengineering, Osaka University, Toyonaka, Osaka, 560-8531, Japan}
%\address[xju]{School of Human Settlement and Civil Engineering, Xi'an Jiaotong University, 
%No.28 Xianning West Road, Xi'an, 710049, China }
%\address[cma]{Center of Numerical Weather Prediction, China 
%                Meteorological Administration, 46 Zhongguancun South street, Beijing, 100081, China}
\cortext[cor]{Corresponding author: Feng Xiao, \it{xiao@es.titech.ac.jp}}

\begin{abstract}
%% Text of abstract
Two variants of the MCV3 scheme are presented based on a flux reconstruction formulation. Different from the original multi-moment constrained finite volume method of third order (MCV3), the multi-moment constraints are imposed at the cell center on the point value, the first and second order derivatives. The continuity of the flux function at cell interfaces are also used as the constraints to ensure the numerical conservation. Compared to the original MCV3 scheme, both two variants have higher numerical accuracy and less restrictive CFL condition for computational stability. Moreover, without the need to solve derivative Riemann problem at cell boundaries, the new schemes benefit the implementations in arbitrary quadrilateral in 2D and hexahedron in 3D.

\end{abstract}


\end{frontmatter}

%%
%% Start line numbering here if you want
%%
%\linenumbers

%% main text

\section{The MCV3 scheme}
We consider the following conservation law  
\begin{equation}
{{\partial u} \over {\partial t}}+{{\partial f} \over {\partial x}} =0,  
\label{1dge}
\end{equation}
where  $u$  is the solution function, and  $f(u)$ the flux function. 
The computational domain is divided into $I$ non-overlapping cells or elements $\Omega_i=[x_{i-{1\over2}},x_{i+{1\over2}}]$, $i=1,2,\cdots,I$. In the MCV3 scheme 3 solution points $x_{ik}$, $k=1,2,3$, are set over  $\Omega_i$ where the solution $u_{ik}$, $k=1,2,3$, is computed.  Suppose that a proper approximation of the flux function  $\hat{f}_i(x)$ is reconstructed, we can immediately update the solutions within  $\Omega_i$ by the following point-wise semi-discretized equations at solution points $x_{ik}$, 
\begin{equation}
{{d u_{ik}} \over {d t}}=-\left [ {{d \hat{f}_i(x)} \over {d x}}\right ]_{ik}, \quad k=1,2,3.  
\label{upu}
\end{equation}

The central task left now is how to reconstruct the flux function $\hat{f}_i(x)$. In principle, the way to reconstruct  $\hat{f}_i(x)$ makes  difference among the numerical schemes\cite{huy07,xiao12}.

For brevity, we make use of a local coordinate system $\xi \in [-1,1]$ that transforms the real mesh cell $x \in [x_{i-1/2},x_{i+1/2}]$ by 
\begin{equation}
 \xi = 2\frac{x-x_{i-\frac{1}{2}}}{\Delta x_i}-1,
\end{equation}
where $\Delta x_i=x_{i+1/2}-x_{i-1/2}$ is the mesh spacing.

The time evolution equations for updating the solutions are correspondingly    
 \begin{equation}
{{d u_{ik}} \over {d t}}=-\left (\frac{d \xi}{dx} \right )_i\left ( {{d \tilde{f}_i(\xi)} \over {d \xi}}\right )_{ik}=-\left (\frac{d \xi}{dx} \right )_i \tilde{f}_{\xi ik}, \quad k=1,2,3.  
\label{upu-local}
\end{equation}
 
We assume that the flux $f(u)$ is a function of solution $u$. The values of the flux at the corresponding points, $f_{ik}$, $k=1,2,3$, are computed directly. The primary reconstruction of flux function is then built by  
\begin{equation}
f_i(\xi)=\sum^3_{k=1}f_{ik}\phi_{k}(\xi).  
\label{lagintp_f3}
\end{equation}
where 
\begin{equation}
\phi_{k}=\prod_{l=1,l\neq k}^3 \frac{\xi-\xi_k}{\xi_l-\xi_k}
\end{equation} 
is the basis function of the Lagrange interpolation. Interpolation \eqref{lagintp_f3} is constructed piecewisely over each cell  element without connection to its neighboring cells, thus cannot be directly used to update the solutions. A modification is required.  

In the MCV3 scheme, the modified flux function is constructed by imposing the continuities of the flux function and its first order derivative at the cell  boundaries, which is realized  by the following constraint conditions, 

\begin{equation}
\left\{
\begin{split}
& \tilde{f}_i(-1)=f^{{\mathcal B}}_i(-1); \\
& \tilde{f}_i(1)=f^{{\mathcal B}}_i(1); \\
& \tilde{f}^{[1]}_{\xi i}(-1)=f^{[1] {\mathcal B}}_{\xi i}(-1); \\
& \tilde{f}^{[1]}_{\xi i}(1)=f^{[1] {\mathcal B}}_{\xi i}(1), 
\end{split} 
\right.
\label{mcv3_1}
\end{equation} 
where 
\begin{equation}
\begin{split}
 & \tilde {f}_{i}(-1)=f^{{\mathcal B}}_{i}(-1)= f^{ {\mathcal B}}_{i-{1\over2}},  \\
 & \tilde {f}_{i}(1)=f^{{\mathcal B}}_{i}(1)= f^{ {\mathcal B}}_{i+{1\over2}}   
\end{split} \label{b-c-f-local}
\end{equation} 
and
\begin{equation}
\begin{split}
 & \tilde {f}^{[1]}_{\xi i}(-1)=f^{[1] {\mathcal B}}_{\xi i}(-1)=\left (\frac{d \xi}{dx} \right )_i^{-1} f^{ [1]  {\mathcal B}}_{xi-{1\over2}},  \\
 & \tilde {f}^{[1]}_{\xi i}(1)=f^{[1] {\mathcal B}}_{\xi i}(1)=\left (\frac{d \xi}{dx} \right )_i^{-1} f^{ [1]  {\mathcal B}}_{xi+{1\over2}}.   
\end{split} \label{b-c-f1-local}
\end{equation} 
The numerical flux at the cell boundary $x_{i-{1\over2}}$, $f^{{\mathcal B}}_{i-{1\over2}}$, and its first order derivative, $f^{ [1]  {\mathcal B}}_{xi-{1\over2}}$ are solved by following Riemann problems(DRP),  
\begin{equation}
\begin{split}
 & f^{{\mathcal B}}_{i-{1\over2}}={\rm Riemann} \left( f^L_{i-{1\over2}}, f^R_{i-{1\over2}} \right), \\
 & f^{ [1]  {\mathcal B}}_{xi-{1\over2}}={\rm DRiemann} \left( f^{[1]L}_{xi-{1\over2}}, f^{[1]R}_{xi-{1\over2}} \right),
\end{split}
\end{equation}
where ``${\rm Riemann}(\cdot \ ,\ \cdot)$" and ``${\rm DRiemann}(\cdot \ ,\ \cdot)$" denote the solvers for the conventional and derivative Riemann problems respectively.

\eqref{mcv3_1} is a Hermite interpolation to determine the modified flux function which is written in a polynomial form as,
\begin{equation}
\left\{
\begin{split}
\tilde{f}_i(\xi)=& \frac{1}{4}\left (f^{{\mathcal B}}_i(-1)-f^{{\mathcal B}}_i(1) +f^{{\mathcal B}}_{\xi i}(-1)+f^{{\mathcal B}}_{\xi i}(1) \right )\xi^3 \\
                 +& \frac{1}{4}\left ( f^{{\mathcal B}}_{\xi i}(1)-f^{{\mathcal B}}_{\xi i}(-1) \right )\xi^2 \\
+ & \frac{1}{4}\left (3f^{{\mathcal B}}_i(1)-3f^{{\mathcal B}}_i(-1) -f^{{\mathcal B}}_{\xi i}(-1)-f^{{\mathcal B}}_{\xi i}(1) \right )\xi \\
+ & \frac{1}{4}\left (2f^{{\mathcal B}}_i(1)+2f^{{\mathcal B}}_i(-1) +f^{{\mathcal B}}_{\xi i}(-1)-f^{{\mathcal B}}_{\xi i}(1) \right ). 
\end{split} 
\right.
\label{mcv3_intp1}
\end{equation}    

The first order derivative (gradient) of \eqref{mcv3_1} reads then, 
\begin{equation}
\left\{
\begin{split}
\tilde{f}_{\xi i}(\xi)=& \frac{3}{4}\left (f^{{\mathcal B}}_i(-1)-f^{{\mathcal B}}_i(1) +f^{{\mathcal B}}_{\xi i}(-1)+f^{{\mathcal B}}_{\xi i}(1) \right )\xi^2 \\
                 +& \frac{1}{2}\left ( f^{{\mathcal B}}_{\xi i}(1)-f^{{\mathcal B}}_{\xi i}(-1) \right )\xi \\
+ & \frac{1}{4}\left (3f^{{\mathcal B}}_i(1)-3f^{{\mathcal B}}_i(-1) -f^{{\mathcal B}}_{\xi i}(-1)-f^{{\mathcal B}}_{\xi i}(1) \right ). 
\end{split} 
\right.
\label{mcv3_intp1-grd}
\end{equation}    

The derivatives of the modified flux function at the solution points are obtained as  
\begin{equation}
\left\{
\begin{split}
& \left ( {{d \tilde{f}_i(\xi)} \over {d \xi}}\right )_{i1}=\tilde{f}_{\xi i1}=\tilde{f}_{\xi i}(\xi_1); \\
& \left ( {{d \tilde{f}_i(\xi)} \over {d \xi}}\right )_{i2}=\tilde{f}_{\xi i2}=\tilde{f}_{\xi i}(\xi_2); \\
& \left ( {{d \tilde{f}_i(\xi)} \over {d \xi}}\right )_{i3}=\tilde{f}_{\xi i3}=\tilde{f}_{\xi i}(\xi_3).  
\end{split} 
\right.
\label{dflux-3pt}
\end{equation} 
The solutions are then immediately computed by \eqref{upu-local} with a proper time integration algorithm. 

In the original MCV3 scheme\cite{ii09}, the solution points are equally spaced and including two cell ends, i.e. $\xi_{1}=-1$, $\xi_{2}=0$ and
$\xi_{3}=1$. The left/right-most solution points coincide with the cell boundaries. In this case, the continuity conditions of flux function at the cell boundaries are automatically satisfied, and only the derivatives of the flux function need to be computed from the DRP. 

 The derivatives of the modified flux function at the solution points are obtained as  
\begin{equation}
\left\{
\begin{split}
& \tilde{f}_{\xi i1}=\tilde{f}_{\xi i}(\xi_1)=f^{{\mathcal B}}_{\xi i}(-1) ; \\
& \tilde{f}_{\xi i2}= \frac{1}{4}\left (3f^{{\mathcal B}}_i(1)-3f^{{\mathcal B}}_i(-1) -f^{{\mathcal B}}_{\xi i}(-1)-f^{{\mathcal B}}_{\xi i}(1) \right ); \\
& \tilde{f}_{\xi i3}=\tilde{f}_{\xi i} (\xi_3)=f^{{\mathcal B}}_{\xi i}(1).  
\end{split} 
\right.
\label{dflux-mcv3_1}
\end{equation} 
It is straightforward to show the following conservation property,  
\begin{equation}
\sum^3_{k=1}\left( \tilde{f}_{\xi ik}\int^1_{-1} \phi_{k}(\xi)d\xi \right)=  \frac{1}{3}\tilde{f}_{\xi i1}+  \frac{4}{3}\tilde{f}_{\xi i2}+ \frac{1}{3}\tilde{f}_{\xi i3} = f^{{\mathcal B}}_{i}(1)-f^{{\mathcal B}}_{i}(-1).
\label{conservation-mcv3-ep}
\end{equation}

The solution points can be chosen as other quadrature point sets, such as the Legendre or Chebyshev Gauss points, but we find from Fourier analysis and numerical tests that the different solution point sets don't alter significantly the numerical result. 

\section{The variants of MCV3 scheme}

We present here two variants by making use of different constraints in determining the modified flux function. 
Instead of the constraint conditions of \eqref{mcv3_1}, we impose the multi-moment constraints at the cell center, 
\begin{equation}
\left\{
\begin{split}
& \tilde{f}_i(-1)=f^{{\mathcal B}}_i(-1); \\
& \tilde{f}_i(1)=f^{{\mathcal B}}_i(1); \\
& \tilde{f}_i(0)=f_i(0); \\
& \tilde{f}^{[1]}_{\xi i}(0)=f^{[1] }_{\xi i}(0); \\
& \tilde{f}^{[2]}_{\xi i}(0)=f^{[2] }_{\xi i}(0). 
\end{split} 
\right.
\label{mcv3_2}
\end{equation}
We retain the continuity of the modified flux function at cell boundaries, which is required from the numerical conservation and stability. The rest of the constraints are determined from the primary interpolation function in terms of derivatives.

\subsection{MCV3 scheme for uniform points with center constraints: MCV3\_UPCC}
Same as in the original MCV3 scheme\cite{ii09}, the solution points are equally spaced and including two cell ends, $\xi_{1}=-1$, $\xi_{2}=0$ and
$\xi_{3}=1$. 

Constraint conditions \eqref{mcv3_2} allows to reconstruct a polynomial of 4th degree, 
\begin{equation}
\left\{
\begin{split}
\tilde{f}_i(\xi)=& \frac{1}{2}\left (f^{{\mathcal B}}_i(-1)+f^{{\mathcal B}}_i(1) -f_{i1}-f_{i3} \right )\xi^4 + \frac{1}{2}\left (f^{{\mathcal B}}_i(1)-f^{{\mathcal B}}_i(-1) +f_{i1}-f_{i3} \right )\xi^3 \\
                 +& \frac{1}{2}\left ( f_{i1}-2f_{i2}+f_{i3} \right )\xi^2 + \frac{1}{2}\left ( f_{i3}-f_{i1} \right )\xi+ f_{i2}. 
\end{split} 
\right.
\label{mcv3_1-intp1}
\end{equation}    
The first-order derivative then reads, 

\begin{equation}
\left\{
\begin{split}
\tilde{f}_{\xi i}(\xi)=& 2\left (f^{{\mathcal B}}_i(-1)+f^{{\mathcal B}}_i(1) -f_{i1}-f_{i3} \right )\xi^3 + \frac{3}{2}\left (f^{{\mathcal B}}_i(1)-f^{{\mathcal B}}_i(-1) +f_{i1}-f_{i3} \right )\xi^2 \\
                 +& \left ( f_{i1}-2f_{i2}+f_{i3} \right )\xi + \frac{1}{2}\left ( f_{i3}-f_{i1} \right ). 
\end{split} 
\right.
\label{mcv3_1-intp1-der}
\end{equation}    

The derivatives of the modified flux function at the solution points are obtained as  
\begin{equation}
\left\{
\begin{split}
& \tilde{f}_{\xi i1}=\tilde{f}_{\xi i}(\xi_1)=2(f_{i1}+f_{i2}-\frac{1}{2}(7f^{{\mathcal B}}_{i}(-1)+f^{{\mathcal B}}_{i}(1)) ; \\
& \tilde{f}_{\xi i2}=\tilde{f}_{\xi i}(\xi_2)= \frac{1}{2}(f_{i3}-f_{i1}); \\
& \tilde{f}_{\xi i3}=\tilde{f}_{\xi i}(\xi_3)=-2(f_{i2}+f_{i3})+\frac{1}{2}(f^{{\mathcal B}}_{i}(-1)+7f^{{\mathcal B}}_{i}(1)).  
\end{split} 
\right.
\label{dflux-mcv3_11}
\end{equation} 

It is straightforward to show the following conservation property,  
\begin{equation}
\sum^3_{k=1}\left( \tilde{f}_{\xi ik}\int^1_{-1} \phi_{k}(\xi)d\xi \right)=  \frac{1}{3}\tilde{f}_{\xi i1}+  \frac{4}{3}\tilde{f}_{\xi i2}+ \frac{1}{3}\tilde{f}_{\xi i3} = f^{{\mathcal B}}_{i}(1)-f^{{\mathcal B}}_{i}(-1).
\label{conservation-mcv3-ep}
\end{equation}

\subsection{MCV3 scheme for Chebyshev points with center constraints: MCV3\_CPCC}

We use the Chebyshev-Gauss points, i.e. $\xi_{1}=-\sqrt{3}/2$, $\xi_{2}=0$ and
$\xi_{3}=\sqrt{3}/2$, as the solution points.
Constraint conditions \eqref{mcv3_2} leads to the following polynomial of 4th degree, 
\begin{equation}
\left\{
\begin{split}
\tilde{f}_i(\xi)=& \frac{1}{6}\left (3f^{{\mathcal B}}_i(-1)+3f^{{\mathcal B}}_i(1) -4f_{i1}+2f_{i2}-4f_{i3} \right )\xi^4 \\
                 &+ \frac{1}{6}\left (3f^{{\mathcal B}}_i(1)-3f^{{\mathcal B}}_i(-1) +2\sqrt{3}f_{i1}-2\sqrt{3}f_{i3} \right )\xi^3 \\
                 +& \frac{1}{6}\left ( 4f_{i1}-8f_{i2}+4f_{i3} \right )\xi^2 + \frac{\sqrt{3}}{3}\left ( f_{i3}-f_{i1} \right )\xi+ f_{i2}. 
\end{split} 
\right.
\label{mcv3_2-intp1}
\end{equation}    
The first-order derivative then reads, 

\begin{equation}
\left\{
\begin{split}
\tilde{f}_i(\xi)=& \frac{2}{3}\left (3f^{{\mathcal B}}_i(-1)+3f^{{\mathcal B}}_i(1) -4f_{i1}+2f_{i2}-4f_{i3} \right )\xi^3 \\
                 &+ \frac{1}{2}\left (3f^{{\mathcal B}}_i(1)-3f^{{\mathcal B}}_i(-1) +2\sqrt{3}f_{i1}-2\sqrt{3}f_{i3} \right )\xi^2 \\
                 +& \frac{1}{3}\left ( 4f_{i1}-8f_{i2}+4f_{i3} \right )\xi + \frac{\sqrt{3}}{3}\left ( f_{i3}-f_{i1} \right ). 
\end{split} 
\right.
\label{mcv3_2-intp1-der}
\end{equation}       

The derivatives of the modified flux function at the solution points are obtained as  
\begin{equation}
\left\{
\begin{split}
& \tilde{f}_{\xi i1}=\tilde{f}_{\xi i}(\xi_1)=\frac{3}{4}\sqrt{3}f_{i1}+\frac{5}{6}\sqrt{3}f_{i2}-\frac{1}{12}\sqrt{3}f_{i3}+\frac{3}{4}\left(\frac{3}{2}-\sqrt{3}\right )f^{{\mathcal B}}_{i}(1)-  \frac{3}{4}\left(\frac{3}{2}+\sqrt{3}\right )f^{{\mathcal B}}_{i}(-1) ; \\
& \tilde{f}_{\xi i2}=\tilde{f}_{\xi i}(\xi_2)= \frac{\sqrt{3}}{3}(f_{i3}-f_{i1}); \\
& \tilde{f}_{\xi i3}=\tilde{f}_{\xi i}(\xi_3)=\frac{1}{12}\sqrt{3}f_{i1}-\frac{5}{6}\sqrt{3}f_{i2}-  \frac{3}{4}\sqrt{3}f_{i3}+\frac{3}{4}\left(\frac{3}{2}+\sqrt{3}\right )f^{{\mathcal B}}_{i}(1)-  \frac{3}{4}\left(\frac{3}{2}-\sqrt{3}\right )f^{{\mathcal B}}_{i}(-1) .  
\end{split} 
\right.
\label{dflux-mcv3_22}
\end{equation}

From \eqref{dflux-mcv3_22}, the numerical conservation can be immediately proved by the following equality,  
\begin{equation}
\sum^3_{k=1}\left( \tilde{f}_{\xi ik}\int^1_{-1} \phi_{k}(\xi)d\xi \right)=  \frac{4}{9}\tilde{f}_{\xi i1}+  \frac{10}{9}\tilde{f}_{\xi i2}+ \frac{4}{9}\tilde{f}_{\xi i3} = f^{{\mathcal B}}_{i}(1)-f^{{\mathcal B}}_{i}(-1).
\label{conservation-mcv3-gp}
\end{equation}

\begin{figure}[h]
\begin{center}
\includegraphics[width=4.5cm]{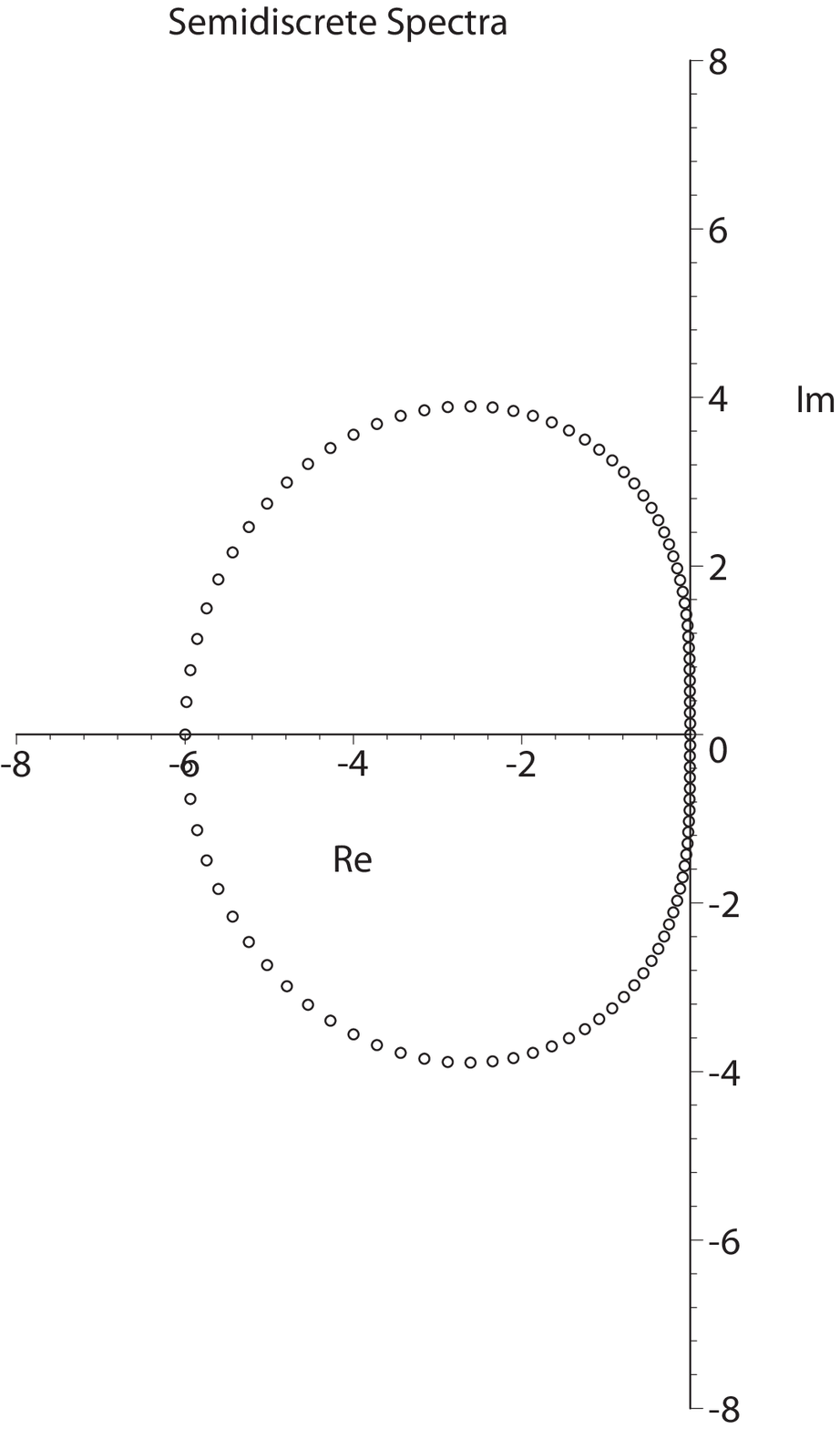} 
  \hspace{0cm} \includegraphics[width=4.5cm]{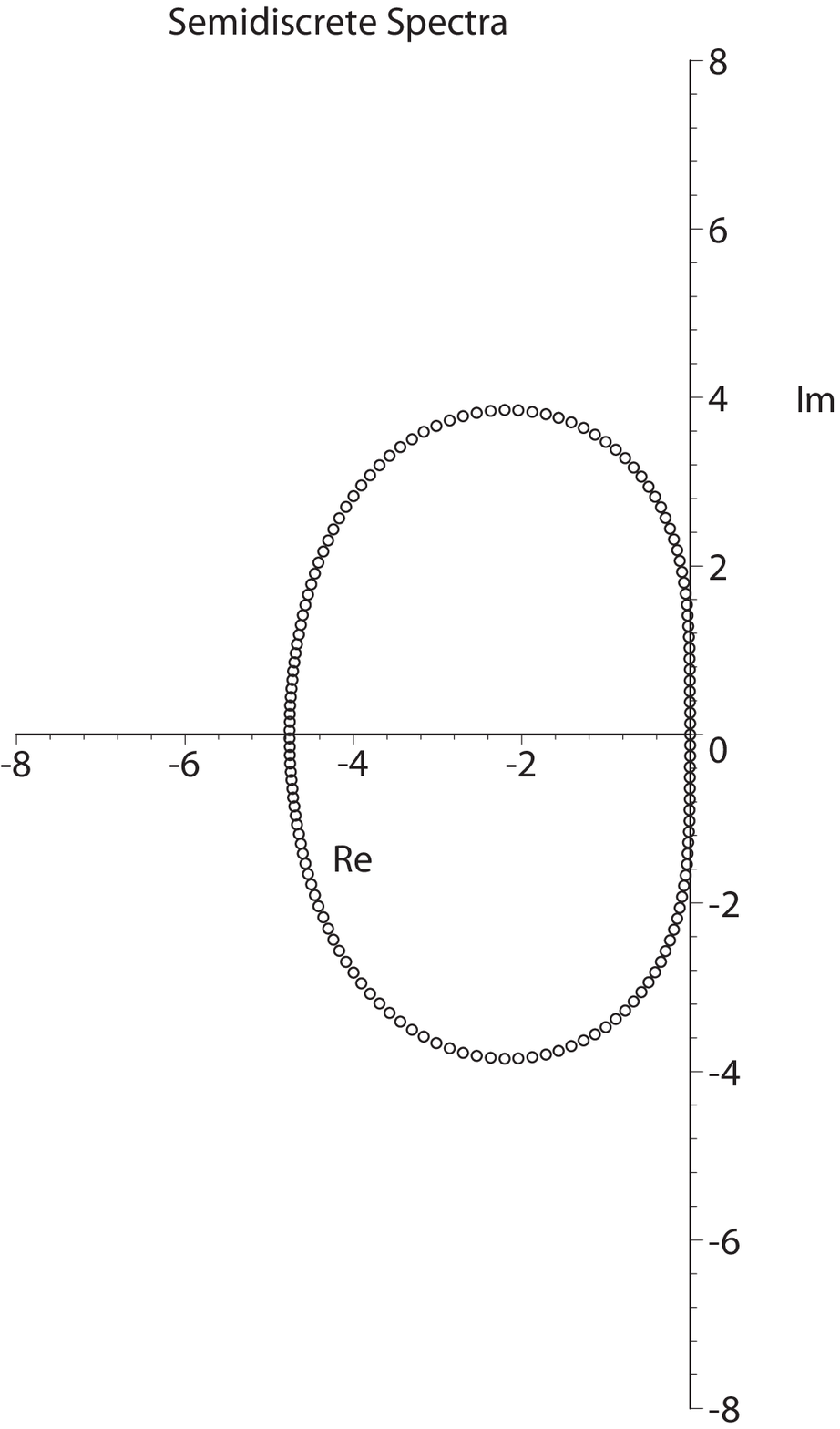} 
   \hspace{0cm} \includegraphics[width=4.5cm]{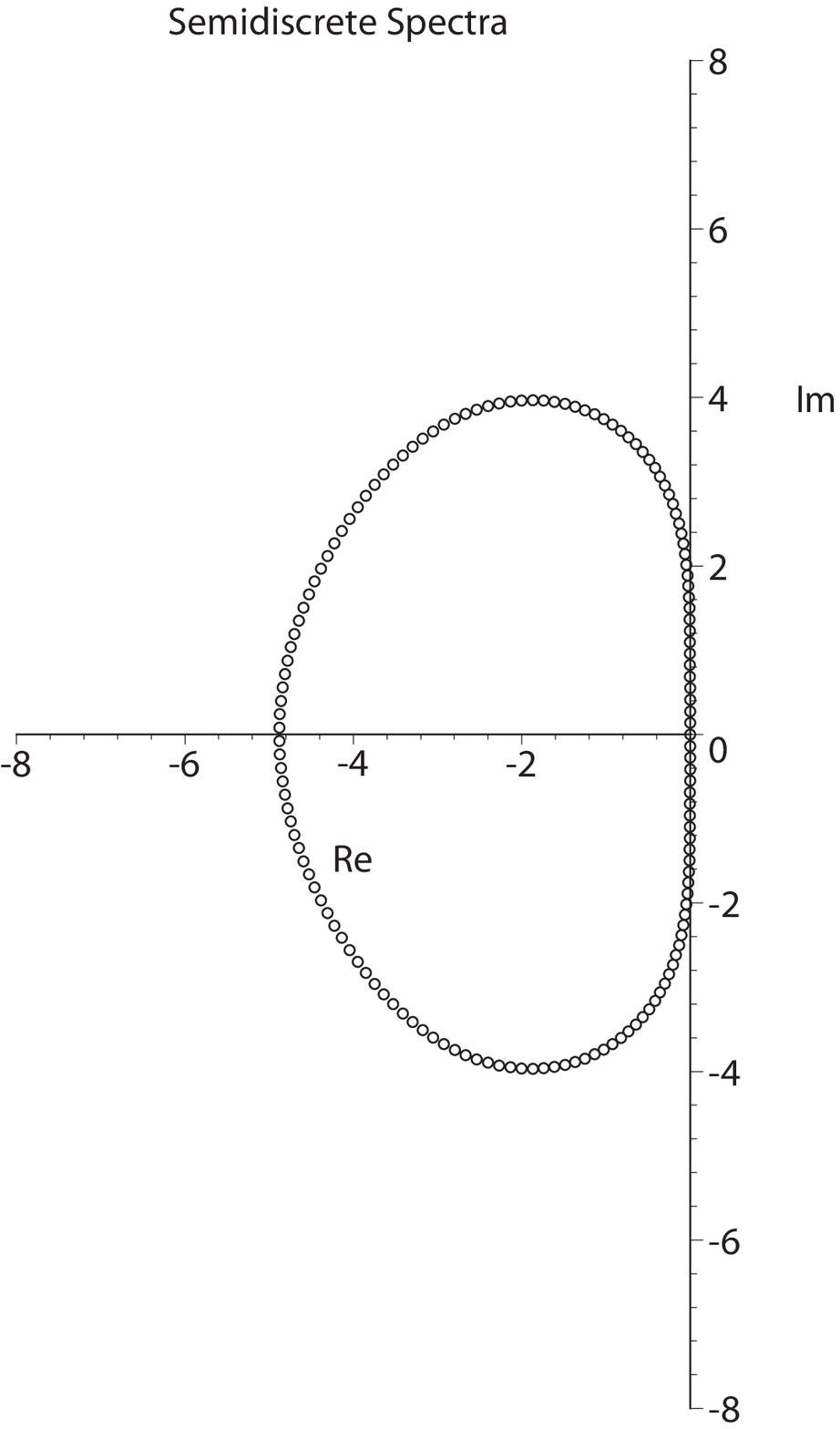} 
\caption{The spectrum of the semi-discrete schemes. Left: MCV3; Center: MCV3\_UPCC, Right: MCV3\_CPCC. }
\label{spec-3} 
\end{center}
\end{figure}

\section{Fourier analysis}

In this section, we evaluated the numerical schemes previously discussed by examining the Fourier mode transported with the following advection equation.
\begin{equation}
{{\partial u} \over {\partial t}}+{{\partial u} \over {\partial x}} =0.  
\label{1dade}
\end{equation} 
Using a wave solution 
\begin{equation}
q(x,t)=e^{I\omega(x+t)},
\label{e-sol}
\end{equation}
and assuming a uniform grid spacing $x_{i+\frac{1}{2}}-x_{i-\frac{1}{2}}=\Delta x$, we have 
$u_{ik}=e^{I\omega(x_i+\xi_l\Delta x/2)}$ and  $u_{(i-1)k}=e^{-I\omega \Delta x} u_{ik}$, which recast the time evolution equations for the solutions into 
\begin{equation}
 \frac{d \mathbf{u}_{i}}{dt}=\mathbf{S}\mathbf{u}_{i}. 
\label{m-form-rv}
\end{equation}

\begin{figure}[h]
\begin{center}
\includegraphics[width=8.5cm]{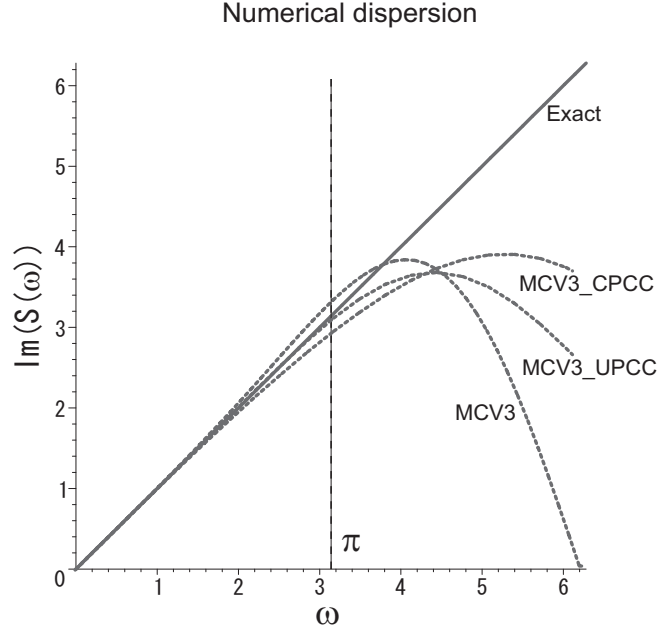}
\caption{Numerical dispersion relations for four-point schemes.  }
\label{disper-3} 
\end{center}
\end{figure}

\begin{figure}[h]
\begin{center}
\includegraphics[width=10.5cm]{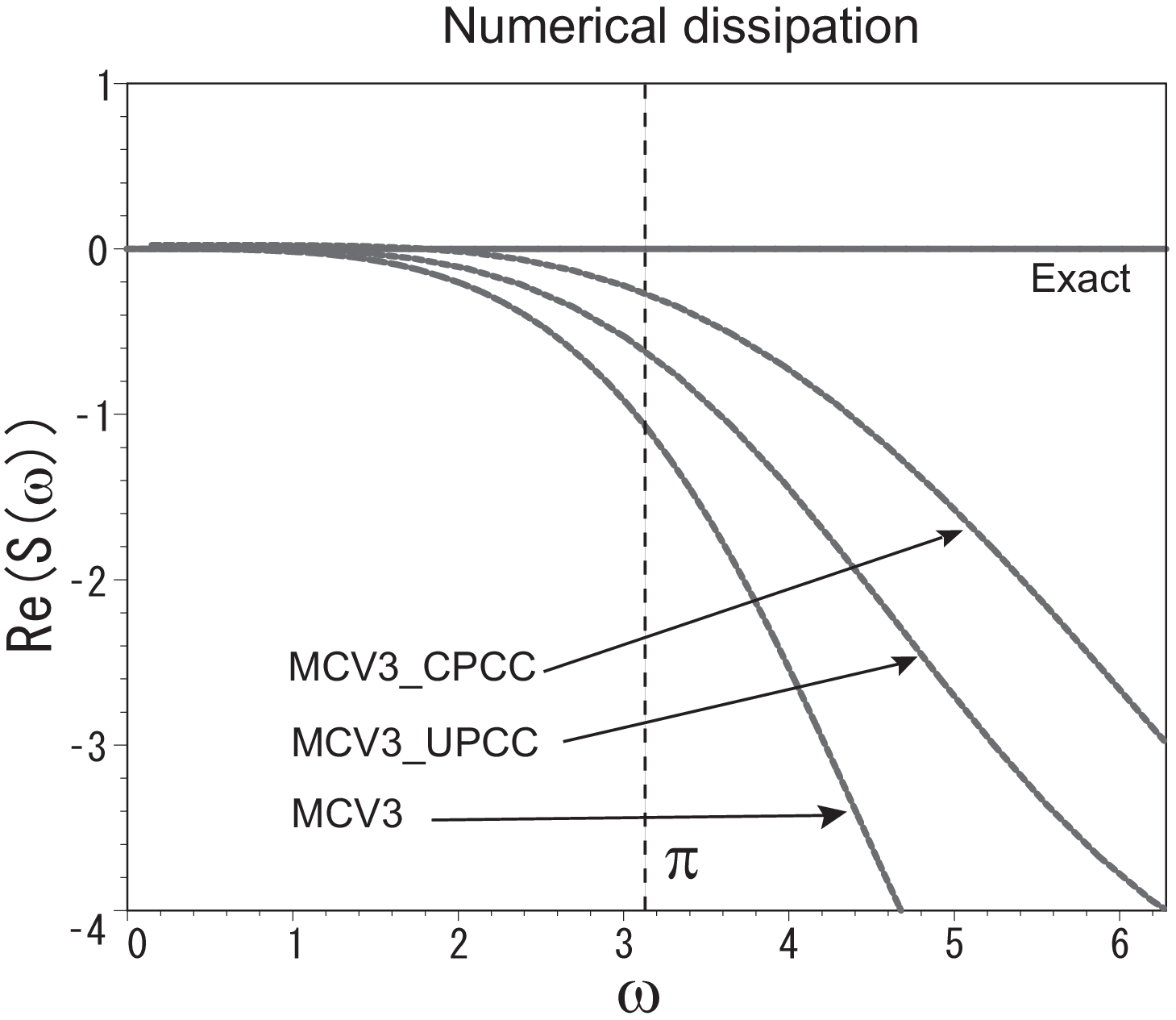} 
\caption{Numerical dispersion relations for three-point schemes. }
\label{gain-3} 
\end{center}
\end{figure}

The properties of the numerical schemes can be examined by analyzing the eigenvalues of \eqref{m-form-rv}. Fig.\ref{spec-3} shows the spectrum (collection of all eigen values) of $\mathbf S$ for different schemes. It is observed that all eigenvalues lie on the left half of the real axis, i.e the negative real parts indicate that all the schemes are stable under the CFL conditions. The allowable CFL numbers for computational stability can be estimated by the largest eigenvalue, the spectral radius $\rho$ for each scheme, i.e. a scheme has a larger spectral radius has to use a smaller CFL number for computational stability. We know from Fourier analysis that $\rho_{\rm MCV3 \_ UPCC}=4.7$,  $\rho_{\rm MCV3 \_ CPCC}=5.4$ and  $\rho_{\rm MCV3}=6.0$, which reveals that  MCV3\_CPCC scheme has the largest stable CFL number. This is confirmed by numerical tests for the linear advection equation. With a 3rd-order Runge Kutta scheme, the largest allowable CFL numbers are 0.47, 0.44 and 0.41 for MCV3\_UPCC, MCV3\_CPCC and MCV3 respectively.

The numerical errors of different schemes can be examined by comparing the principal eigenvalue of  $\mathbf S$, $\lambda^p_{\mathbf S(\omega \Delta x)}$, with the exact solution, $-I\omega$, of the advection equation \eqref{1dade} for initial condition,  
\begin{equation}
q(x,0)=e^{I\omega x}. 
\label{ic}
\end{equation}
The error of a given semi-discrete formulation is 
\begin{equation}
E(\omega)=\lambda^p_{\mathbf S(\omega \Delta x)}-I\omega, 
\label{err-def}
\end{equation}
and the convergence rate is evaluated by 
\begin{equation}
m=\left(\ln\left(\frac{E(\omega \Delta x)}{E(\omega \Delta x/2)}\right)/\ln(2)\right )-1. 
\label{err-def}
\end{equation}

\begin{table}[h]
\begin{center}\label{error-3ps}
\caption{Numerical errors and convergence rates.}
\begin{tabular}{lcccc} \hline
 Scheme & &$\omega \Delta x=\pi/8$ & $\omega \Delta x=\pi/16$ & order \\ \hline 
 MCV3 & &$-3.25\times 10^{-4}-3.33\times 10^{-5}i$  & $-2.06\times 10^{-5}-1.07\times 10^{-6}i$ & 2.99 \vspace{0.1cm} \\
 MCV3\_UPCC & &$-1.65\times 10^{-4}-2.22\times 10^{-6}i$  & $-1.03\times 10^{-5}-6.81\times 10^{-9}i$ & 3.00 \vspace{0.1cm} \\
 MCV3\_CPCC & &$-3.15\times 10^{-6}+1.91\times 10^{-5}i$  & $-4.93\times 10^{-8}+6.05\times 10^{-7}i$ & 3.99 \vspace{0.1cm} \\ \hline
\end{tabular}                            
\end{center}
\end{table}

The numerical errors of the three schemes are given in Table 1. MCV3\_UPCC and MCV3\_CPCC are more accurate than the original MCV3 scheme. Similar to the original MCV3, MCV3\_UPCC shows a 3rd-order convergence rate, while MCV3\_CPCC has a 4th-order convergence rate. 

The  dispersion and dissipation relations of the spatial discretization can be evaluated by plotting the real and imaginary parts of the principal eigenvalues as a functions of the wave number $\omega$. From Fig.\ref{disper-3}, we find that the MCV3\_UPCC has the most accurate numerical dispersion. MCV3\_CPCC has a different dispersion behavior compared to MCV3 and MCV3\_UPCC. The numerical dissipations  are plotted in Fig.\ref{gain-3}.  MCV3\_CPCC is the best, which is also observed in numerical tests. Both MCV3\_UPCC and MCV3\_CPCC have improved dissipation accuracy over the original MCV3 scheme.

We further give the Taylor expansion of the eigen values of  $\mathbf S$ with respect to the mesh size in Table 2. 
\begin{table}[h]
\begin{center}\label{eigen-taylor-3ps}
\caption{Taylor expansion of the eigenvalues in terms of the mesh size. The principal eigenvalue $\lambda^p_{\mathbf S(\omega \Delta x)}$ is $\lambda_1$ for all schemes. }
\begin{tabular}{lll} \hline
 Scheme & & Eigen values \\ \hline  \hline
 MCV3 & &$\lambda_1=-I\omega -1.39\times 10^{-2}\omega^4 \Delta x^3 -3.70\times 10^{-3}I\omega^5 \Delta x^4+O(\Delta x^5) $   \\
 & &$\lambda_2=-6/\Delta x +3I\omega +\omega^2\Delta x $   \\
 & &$\lambda_3=0 $   \\
 \hline
  MCV3\_UPCC & &$\lambda_1=-I\omega -6.94\times 10^{-3}\omega^4 \Delta x^3 -2.31\times 10^{-4}I\omega^5 \Delta x^4+O(\Delta x^5) $   \\
 & &$\lambda_2=(-4+2.83I)/\Delta x +(0.71+I)\omega +(0.25-8.84\time 10^2I)\omega^2\Delta x + O(\Delta x^2)$   \\
 & &$\lambda_3=(-4-2.83I)/\Delta x +(-0.71+I)\omega +(0.25+8.84\time 10^2I)\omega^2\Delta x + O(\Delta x^2)$   \\
 \hline
   MCV3\_CPCC & &$\lambda_1=-I\omega +2.08\times 10^{-3}I\omega^5 \Delta x^4 -8.68\times 10^{-4}\omega^6 \Delta x^5+O(\Delta x^6) $   \\
 & &$\lambda_2=(-3+3.87I)/\Delta x +(1.16+0.5I)\omega -0.31I\omega^2\Delta x + O(\Delta x^2)$   \\
 & &$\lambda_3=(-3-3.87I)/\Delta x +(-1.16+0.5I)\omega +0.31I\omega^2\Delta x + O(\Delta x^2)$    \\
 \hline  

\end{tabular}                            
\end{center}
\end{table}
 
Consistent with the observations aforementioned,  we find from  Table 2 that  MCV3\_UPCC and  MCV3\_CPCC have less truncation errors in both dissipation (real part) and dispersion (imaginary part) compared to the original MCV3. MCV3 and MCV3\_UPCC have a third order accuracy in dissipation and all schemes have a fourth order accuracy in dispersion. The  MCV3\_CPCC is superior in dissipation accuracy which is fifth order, two orders higher than the others. The computational modes are represented by $\lambda_2$ and  $\lambda_3$. It is observed that all the schemes have real negative parts with the leading terms  of order $O(\Delta x^{-1})$ in  $\lambda_2$ and  $\lambda_3$, which means that the computational modes will be exponentially dampened out.  

\section{A few remarks}
\begin{itemize}
\item 
The proposed variants have improved numerical features in both numerical accuracy and computational efficiency compared to the original MCV3 scheme. 
\item
In the new schemes, only the continuity of flux function is required at the cell boundaries where the constraint on the derivative of flux is not required anymore. This makes the schemes directly applicable to any quadrilateral and xahedral mesh. 
\item
Schemes with more solution points and higher order accuracy can be devised by the same spirit. 
\end{itemize}

%%%%%%%%%%%%%%%%%%%%%%%%
% References
%%%%%%%%%%%%%%%%%%%%%%%%

\end{document}